# Pattern engineering of living bacterial colonies using meniscus-driven fluidic channels

Vasily Kantsler[1,2,*], Elena Ontañón-McDonald[3,#], Cansu Kuey[3,#], Manjari J Ghanshyam[3], Maria Chiara Roffin[1] and Munehiro Asally[3,4,*]

[1] Department of Physics, University of Warwick, Coventry, CV4 7AL, United Kingdom
[2] Warwick Medical School, University of Warwick, Coventry, CV4 7AL, United Kingdom
[3] School of Life Sciences, University of Warwick, Coventry, CV4 7AL, United Kingdom
[4] Warwick Integrative Synthetic Biology Centre, University of Warwick, Coventry, CV4 7AL, United Kingdom
[#] Equal contribution
[*] Correspondence should be addressed to VK (v.kantsler@warwick.ac.uk) or MA (m.asally@warwick.ac.uk)

Engineering spatially organized biofilms for creating adaptive and sustainable biomaterials is a forthcoming mission of synthetic biology. Existing technologies of patterning biofilm materials suffer limitations associated with the high technical barrier and the requirements of special equipment. Here we present controlled meniscus-driven fluidics, MeniFluidics; an easily implementable technique for patterning living bacterial populations. We demonstrate multiscale patterning of living-colony and biofilm formation with submillimetre resolution. Relying on fast bacterial spreading in liquid channels, MeniFluidics allows controlled anisotropic bacterial colonies expansion both in space and time. The technique has also been applied for studying collective phenomena in confined bacterial swarming and organizing different fluorescently labelled *Bacillus subtilis* strains into a converged pattern. We believe that the robustness and low technical barrier of MeniFluidics offer a tool for developing living functional materials, bioengineering and bio-art, and adding to fundamental research of microbial interactions.

**KEYWORDS**

## INTRODUCTION

Owing to their multifunctional nature, bacterial communities represent sources of inspirations and opportunities for engineers and scientists to create adaptive and programmable living functional materials (*1–3*). Self-organized bacterial communities, such as biofilms and swarms, exhibit multicellular behaviours and promote the survival of cells in harsh environments (*4–7*). Through metabolic and biophysical interactions, biofilms produce valuable chemicals that can be used for next-generation batteries, biodegradable plastics and biomedical applications (*8*, *9*). For these collective functions and interactions, the arrangements of microbes in space and time are crucial (*10*, *11*). Bacteria also serve as a model system to study active fluids (*12*). Recent progress in theoretical understanding of topological active matter (*13–15*) raised the possibility of creating dynamic and adaptive metamaterials (*16*), which requires advances in experimental techniques for controlling active constituents and their collectives on multiple scales. Therefore, controlling the spatial distributions of microbes has implications for the fundamental research into engineering, biophysics, microbiology and ecology and holds a promise in material science, environmental restoration and bio-productions.

Substantial amount of bioengineering effort in the last decade has been exploited for developing living cells and biomaterials that self-organize into desired patterns (*17*, *18*). For example, programable nano-objects and biomaterials were developed by functionalizing amyloid fibril, the main protein component of biofilm matrix (*19*, *20*). Bacterial populations can be organized into patterns using inkjet printers (*21*), 3D printing (*22*), biofilm lithography (*23*) or genetic engineering (*24–26*). Yet, controlling the active matter and collective dynamics of living cells largely remains as a challenge. Furthermore, these existing methods are not widely used because of the associated high-technical barriers that demand precise control of tools, customized equipment and/or extensive genetic modifications. A robust and easy method to precisely control microbial patterns in space and time could unleash the full potential of biofilm engineering and facilitate biomaterial innovations.

This study presents a technique for micro-patterning fluid meniscus layers, hereby named as MeniFluidics. We demonstrate that local acceleration of bacterial spreading via MeniFluidics can shape biofilms into desired patterns, arrange multiple stains into controlled proximity and provide a tool to study self-organisation dynamics of active fluids in confinement.

## RESULTS

**Rationalized gel surface structuring for micropatterned channel formation**

Our approach is based on the notion that a structured agar gel with a sharp concave edge is wetted by a fluid meniscus layer (Fig. 1, S1 and S2). The observed meniscus radii of curvature are much smaller than the water capillary length, which is defined by the balance of the fluid surface tension to the gravity force: $L_c = \sqrt{\sigma/\rho g}$, where $\sigma$ and $\rho$ are the surface tension and the density of the fluid, respectively. The observed small meniscus radii can be attributed to the reduced water pressure in agar ($P_{agar} < P_{atm}$) due to elastic contraction of the gel. After gelation, the elastic matrix of agar contracts due to evaporation of the water phase. This creates an extensile force acting on the fluid, which gives rise to a reduced pressure in the agar. When we assume a linear response of the elastic agar gel, the pressure $P_{agar}$ can be estimated as $P_{atm} - E\Delta$, where $\Delta = \delta/A$ is a ratio of the evaporated volume of water ($\delta$) from the agar to the total volume of agar layer ($A$) after gelation, and $E$ is the Young's elastic modulus of the agar gel. The meniscus radii, $R_1$, can be estimated by the Young-Laplace equation:

$$E\Delta = \sigma \left(\frac{1}{R_1} + \frac{1}{R_2}\right) \quad (1)$$

where $R_1$ and $R_2$ are principal radii of the curvature. In the case of a plane meniscus (Fig. 1A-C, S1A and B), $1/R_2$ is negligible and we have $R_1 = \frac{\sigma}{E}\Delta^{-1}$. Measured variation of $R_1$ as a function of $\Delta$ for 1.5% agar gel in water reveals that the decay is slightly slower than $\Delta^{-1}$ (Fig. S1). This indicates that agar does not act as an ideal Hookean spring, but possibly exhibits plastic deformations. The measurements estimated the Young modulus of the agar to be $\sim 10^4 - 10^5 Pa$, which is in agreement with reported values (27). For $\Delta$ in a range of a few percent, we obtain the meniscus radius of curvature $R_1$ to be in the order of a few tens of microns, accounting also for the fact that a complex medium reduces the surface tension compare to water (28). Hence, structuring agar surface should create microns-size meniscus liquid layers that provide a favourable path for bacteria to spread. Such fluidic channels, with a cross section size $h \approx 0.3R_1$, can be used for spatial control over biofilm patterning.

**Controlled biofilm spreading by MeniFluidics**

To implement the above-described approach and create liquid meniscus micro channels, we structured an agar surface by placing a micropatterned PDMS mould on an agar plate during solidification (Fig. 1B, C, *see also* Fig. S3 and S4). When fluorescently labelled *B. subtilis* cells were inoculated to the agar, random motions of cells were observed near, i.e. within ~16 μm, from the edge of the micropatterned structures (Fig. 1D, Fig. S5 and Movie S1). This observation confirms the formation of meniscus layers along the structured agar surfaces. To examine whether the meniscus channels can accelerate the spreading of *B. subtilis* colonies, bacteria were inoculated on the flat agar surface approximately 3 mm away from a micropatterned structure (Fig. 1E). The biofilm followed the meniscus pattern as a result of accelerated spreading of the bacterial cells along MeniFluidics (Fig. 1F, G, and Movie S2). The meniscus-accelerated expansion was also confirmed with *E. coli* (Fig. S6).

The gained spatial control over biofilm expansion posed a question of whether the overall expansion speed can be tuned. To this end, we considered geometrically imposed control over the meniscus radii by introducing a convex structure that creates the second principle radius of curvature, $R_2$, for the meniscus (Fig. S2). As denoted by the Young-Laplace equation (Eq. 1), introduction of the second principal radius of curvature $R_2$ should alter $R_1$, thus, the MeniFluidics channel size. When $E\Delta$ and $\sigma$ are constant, a small negative convex $R_2$ leads to smaller $R_1$, while small positive concave $R_2$ increases $R_1$. Accordingly, the bacteria translocation along MeniFluidic channels can cease at a small radius of curved convex structure when $h$ decreases to micron size. This suggests that highly curved structure supporting the meniscus should introduce breaks in MeniFluidics.

To test this conjecture, we introduced hinges, which can break the bacterial flow at their high curvature tips (Fig. 2A). More specifically, we designed the structures with periodically spaced hinges with a period $D_g$. This design made the biofilm expansion along MeniFluidics to be periodically halted (Fig. 2B, C, and Movie S3). The distances between halts were in good agreement with the distance between hinges ($D_g$), indicating that the hinge structures slowed down the expansion (Fig. 2B). The overall velocities of biofilm expansion were determined by linear regression of colony edge positions over time (Fig. S7). Examining MeniFluidics with various $D_g$, we find that the overall velocity is proportional to $D_g$ (Fig. 2D). Overall, the above results indicate that MeniFluidics enables controlling the biofilm expansion both spatially and temporally.

**Pattern engineering of biofilms and swarming dynamics**

To demonstrate the versatility of MeniFluidics, we designed and implemented arbitrary patterns, such as a snowflake-like pattern (Fig. 3A). Bacteria were inoculated at the centre of the design and biofilm growth was monitored by time-lapse imaging. Biofilm was formed following the MeniFluidic channels and emerged into the designed shape (Fig. S8 and Movie S4). Furthermore, structured agar provided a unique platform for biophysics studies on bacterial swarming and active turbulence. We implemented MeniFluidics structure to confine the swarming colony in micron-sized domains consisting of circular, elliptical, and rectangular wells. The structured domains were populated by swarming bacteria, which facilitated bacterial turbulence phenomenon under confinement (Fig. 3B-F and Movie S5). The results revealed an interesting case of spiral vortices, vortex lattices and bacterial turbulence which are reminiscent of what have been reported in microfluidic confinement and isolated droplets (*29–31*). An advantage of MeniFluidics here is that it extends the life-time of collective dynamics in the system due to bacteria's unrestricted access to oxygen in contrast to the microfluidics experiments, which enables to implement the method to develop and study biological metamaterials.

The spatio-temporal control by MeniFluidics could also allow organizing different bacterial species and strains into defined spatial patterns. As a proof of concept, we designed an interdigitated pattern in which two bacterial strains can form a stripe pattern in the middle (Fig. 4A). For visualization of different strains, we used fluorescently labelled bacterial strains and inoculated them at the other ends of the pattern. Specifically, we used *B. subtilis* strains expressing yellow fluorescence protein (YFP) or cyan fluorescence protein (CFP). YFP-labelled and CFP-labelled strains were organized into the designed pattern and formed an alternating pattern in the middle where the distance between two strains is precisely controlled (Fig. 4B, C). A notable feature of MeniFluidics is that it does not require any special equipment or use of genetically modified strains. Even if microfabricated moulds are not readily available, the method can be applied because meniscus layers, and thus MeniFluidics, can be formed by any boundary of an object placed on an agar surface. As an example, we examined and confirmed that a metal piece placed on agar surface accelerated the expansion of a biofilm (Movie S6).

**DISCUSSION**

We presented a robust method of patterning bacterial populations in space and time. An exciting application of the method, among many others, could be development of living materials for bioremediation via spatial control of microbial communities. For example, MeniFluidics could be designed to facilitate bacteria with biodegradation capability to spread effectivity in a contaminated environment, where biodegradation of toxic compounds requires spatial organization of multiple engineered strains (*32*). The interdigitated pattern presented in this work could be useful to optimize the distances between two engineered strains for the efficiency. It can also be used to induce autoinhibition of bacterial population via sibling competition mechanism (*33*, *34*). Another application could be using the fast boundary propagation of bacteria along the MeniFluidic pattern for a quick assessment of antimicrobial resistance. The antimicrobial resistance agar plate can be modified by parallel ridges structure, while asymmetry in propagation of the colony along the menisci can indicate antimicrobial response. Furthermore, the proof of principle of the method may be extended for tissue engineering with mammalian cells because mammalian cells can only thrive in liquid, such as in MeniFluidic channels.

Our observation is in an interesting conjunction with the phenomenon known as "fungal highways" where fungi hypae and mycelium promote the dispersal of bacteria on cheese, potatoes and various surfaces (*35*). While the exact physical mechanism behind the fungal highways is unknown, our study suggests that fungal hypae may create natural MeniFluidics in which bacteria spreading is accelerated. Fungal highways have been observed with various species of fungi and bacteria, which supports the idea that physics underlines this cross-kingdom interaction. MeniFluidics could allow dissecting the biochemical interactions from physical interactions between fungi and bacteria through controlled experimentation of multispecies microbiota. Therefore, this method is not only useful for engineering living materials but also to advance experimental research in microbial biophysics and ecology.

In summary, we demonstrate that MeniFluidics offers a unique method to control dynamics and spatio-temporal organisation of the living matter such as bacteria swarms and biofilms at multiple scales, which can be easily adopted to a plethora of fundamental studies and applications.

**METHODS**

**Strains and growth conditions**

*B. subtilis* and *E. coli* cells were routinely grown in Lysogeny Broth (LB) or on LB agar plate (agar 1.5% (w/v), agar 1% (w/v) for swarming experiments). The strains used in this study are listed in Table 1. For MeniFluidics experiments, a single colony from LB agar plate was inoculated into LB and cultured for ~6 hours at 37°C with aeration (200 rpm). Cells were then inoculated onto minimal salts glutamate glycerol (MSgg) media (*4*): 5 mM potassium phosphate (pH 7.0), 100 mM morpholine propane sulfonic acid (MOPS) (pH 7.0), 2 mM $MgCl_2$, 700 µM $CaCl_2$, 50 µM $MnCl_2$, 100 µM $FeCl_3$, 1 µM $ZnCl_2$, 2 µM thiamine-HCl, 0.5% (v/v) glycerol, 0.5% (w/v) monosodium glutamate and 1.5% (w/v) agar. For swarming experiments, 5 µl of the culture was inoculated on 1% (w/v) agar plate 8 hours prior to the swarming measurements.

**PDMS stamps and patterning agar surfaces**

Patterns for MeniFluidics were fabricated on four-inch silicon wafers using a conventional soft lithography technique (*see* Fig. S3). The patterns were transferred to Polydimethylsiloxane (PDMS) (Sylgard 184, Dow Corning Corp) using soft lithography technique. PDMS prepolymer was mixed at a ratio of 10:1 (w/w) with curing agent and poured on top of silicon wafer mould. After degassing the mixture, PDMS was cured at 100°C for 1 hour. The cured PDMS was peeled off from Silicon wafer mould for stamping patterns on agar plates. To transfer the features onto agar plates, MSgg or LB media were poured into plastic dishes (40 mL to 150 mm plates, 20 mL to 90 mm plates) and set to solidify for 30-60 min. Solidified plates were warmed on a 70°C plate while 10 ml (for 150 mm petri dish) and 5 ml (for 90 mm petri dish) media were poured as a second layer. Autoclaved PDMS with features, prewarmed at 50°C, were gently placed on the second layer. Bubbles were removed by tapping with the end of the forceps. After solidification of the second layers, PDMS was removed. Plates were dried at 37°C overnight.

**Time-lapse imaging**

Biofilms formation was recorded using a digital single-lens reflex camera (DSLR, Nikon D5300)

in a 30°C incubator. AF-S DX Micro NIKKOR 40mm f/2.8G macro lens and AF-P DX NIKKOR 10-20mm f/4.5-5.6G VR ultra-wide zoom lens were used. A custom-built imaging platform was assembled to place the camera perpendicular to plates (Fig. S4). The items used for the assembly of the platform are listed in Table 2. Images were taken every hour using the built-in internal time shooting function of D5300. For microscopy, images were obtained by an inverted fluorescence microscope (Leica, DMi8) with HC PL FLUOTAR 40x/0.60 Ph2.

**Video-microscopy of swarming and menisci radii**

The swarming under confinement and the measurements of the menisci radii was recorder by inverted microscope Nikon TE2000U with 20x/0.45 Ph1 lens at 50fps using Point Grey GS3-U3-23S6M-C camera (FLIR). The velocity fields from the image sequences were obtained via open source MatPIV toolbox (version 1.61). The menisci radii measurements were undertaken with 160 microns structure on 1.5% agar gel in deionised water. The agar plate has been weighted immediately after gel moulding. The relative reduced mass of the gel, $\Delta$, has been obtained through the mass measurements with a precision in $\Delta$ of $0.6 \times 10^{-3}$. The image was aligned with the inner corner of the structure, then the focus had been moved to the tip of the meniscus. The intensity profile has been averaged along the image symmetry (y-axis) and the tip position has been identified manually with an uncertainty of 0.9 microns (see Fig. S1).

**ACKNOWLEDGEMENTS**

We acknowledge Yuandi Wei and Dario Bazzoli for their experimental attempts with MeniFluidics. We thank Drs. Marco Polin, Orkun Soyer, Jorn Dunkel and Dong-yeon Lee for their critical comments to early versions of the manuscript. We also thank Drs. Petr Denissenko and Gareth Alexander for helpful discussions. VK gratefully acknowledges support from British Council Link Travel Grant (2017-RLTG8-10229) and welcoming hosting of Dr. Avraham Beer in his lab at Jacob Blaustein Institute for Desert Research, Israel. This project was financially supported by University of Warwick School of Life Sciences pump prime funding and BBSRC/EPSRC Synthetic Biology Centre grant (BB/M017982/1) to MA and EPSRC/BBSRC Synthetic Biology Centre for Doctoral Training grant (EP/L016494/1) to CK.

**AUTHOR CONTRIBUTIONS**

VK and MA conceived the project and designed experiments. VK analyzed the theory of meniscus formation. EOM, CK and VK performed experiments. EOM designed bio-art patterns. VK and MJG prepared materials for experiments. VK, MA and CK analyzed data. MA and VK supervised the project. MA and VK wrote the manuscript with the inputs from other authors.

# REFERENCES

1. M. Schaffner, P. A. Rühs, F. Coulter, S. Kilcher, A. R. Studart, 3D printing of bacteria into functional complex materials. *Sci. Adv.* **3** (2017), doi:10.1126/sciadv.aao6804.

2. P. Q. Nguyen, Synthetic biology engineering of biofilms as nanomaterials factories. *Biochem. Soc. Trans.* **45**, 585–597 (2017).

3. K. Brenner, L. You, F. H. Arnold, Engineering microbial consortia: a new frontier in synthetic biology. *Trends Biotechnol.* **26**, 483–489 (2008).

4. M. Asally, M. Kittisopikul, P. Rue, Y. Du, Z. Hu, T. Cagatay, A. B. Robinson, H. Lu, J. Garcia-Ojalvo, G. M. Suel, Localized cell death focuses mechanical forces during 3D patterning in a biofilm. *Proc. Natl. Acad. Sci.* . **109**, 18891–6 (2012).

5. J. Liu, A. Prindle, J. Humphries, M. Gabalda-Sagarra, M. Asally, D. D. Y. D. Y. D. D. Lee, S. Ly, J. Garcia-Ojalvo, G. M. G. M. Süel, Metabolic co-dependence gives rise to collective oscillations within biofilms. *Nature*. **523**, 550–4 (2015).

6. C. D. Nadell, K. Drescher, K. R. Foster, Spatial structure, cooperation and competition in biofilms. *Nat. Rev. Microbiol.* **14**, 589–600 (2016).

7. J. N. Wilking, T. E. Angelini, A. Seminara, M. P. Brenner, D. a. Weitz, Biofilms as complex fluids. *MRS Bull.* **36**, 385–391 (2011).

8. T. K. Wood, S. H. Hong, Q. Ma, Engineering biofilm formation and dispersal. *Trends Biotechnol.* **29**, 87–94 (2011).

9. S. Douglas, Mineral formation by bacteria in natural microbial communities. *FEMS Microbiol. Ecol.* **26**, 79–88 (2002).

10. X. Jiang, C. Zerfaß, S. Feng, R. Eichmann, M. Asally, P. Schäfer, O. S. Soyer, Impact of spatial organization on a novel auxotrophic interaction among soil microbes. *ISME J.* **12** (2018), doi:10.1038/s41396-018-0095-z.

11. K. Nagy, Á. Ábrahám, J. E. Keymer, P. Galajda, Application of microfluidics in experimental ecology: The importance of being spatial. *Front. Microbiol.* **9** (2018), doi:10.3389/fmicb.2018.00496.

12. H. H. Wensink, J. Dunkel, S. Heidenreich, K. Drescher, R. E. Goldstein, H. Löwen, J. M. Yeomans, Meso-scale turbulence in living fluids. *Proc. Natl. Acad. Sci.* . **109**, 14308–14313 (2012).

13. A. Souslov, B. C. Van Zuiden, D. Bartolo, V. Vitelli, Topological sound in active-liquid metamaterials. *Nat. Phys.* **13**, 1091–1094 (2017).

14. F. G. Woodhouse, J. Dunkel, Active matter logic for autonomous microfluidics. *Nat. Commun.* **8**, 1–7 (2017).

15. K. Dasbiswas, K. K. Mandadapu, S. Vaikuntanathan, Topological localization in out-of-equilibrium dissipative systems. *Proc. Natl. Acad. Sci.* . **115**, E9031–E9040 (2018).

16. A. Alù, Prime time. *Nat. Mater.* **15**, 1229–1231 (2016).

17. A. Y. Chen, C. Zhong, T. K. Lu, Engineering living functional materials. *ACS Synth. Biol.* **4**, 8–11 (2015).

18. C. Gilbert, T. Ellis, Biological Engineered Living Materials: Growing Functional Materials with Genetically Programmable Properties. *ACS Synth. Biol.* **8**, 1–15 (2018).

19. W. H. Depas, M. R. Chapman, Microbial Manipulation of the Amyloid Fold. *Res. Microbiol.* (2012), doi:10.1016/j.resmic.2012.10.009.

20. A. Taglialegna, I. Lasa, J. Valle, Amyloid Structures as Biofilm Matrix Scaffolds. *J. Bacteriol.* **198**, 2579–2588 (2016).

21. J. Merrin, S. Leibler, J. S. Chuang, Printing multistrain bacterial patterns with a piezoelectric inkjet printer. *PLoS One*. **2** (2007), doi:10.1371/journal.pone.0000663.

22. J. L. Connell, E. T. Ritschdorff, M. Whiteley, J. B. Shear, 3D printing of microscopic bacterial communities. *Proc. Natl. Acad. Sci.* . **110**, 18380–5 (2013).

23. X. Jin, I. H. Riedel-Kruse, Biofilm Lithography enables high-resolution cell patterning via optogenetic adhesin expression. *Proc. Natl. Acad. Sci.* **115**, 3698–3703 (2018).

24. D. Karig, K. M. Martini, T. Lu, N. A. DeLateur, N. Goldenfeld, R. Weiss, Stochastic Turing patterns in a synthetic bacterial population. *Proc. Natl. Acad. Sci.* **115**, 6572–6577 (2018).

25. M. J. Liao, M. O. Din, L. Tsimring, J. Hasty, Rock-paper-scissors: Engineered population dynamics increase genetic stability. *Science*. **365**, 1045–1049 (2019).

26. A. Kan, I. Del Valle, T. Rudge, F. Federici, J. Haseloff, Intercellular adhesion promotes clonal mixing in growing bacterial populations. *J. R. Soc. Interface*. **15** (2018), doi:10.1098/rsif.2018.0406.

27. V. T. Nayar, J. D. Weiland, C. S. Nelson, A. M. Hodge, Elastic and viscoelastic characterization of agar. *J. Mech. Behav. Biomed. Mater.* **7**, 60–68 (2012).

28. P. A. Rühs, L. Böni, G. G. Fuller, R. F. Inglis, P. Fischer, In-situ quantification of the interfacial rheological response of bacterial biofilms to environmental stimuli. *PLoS One*. **8** (2013), doi:10.1371/journal.pone.0078524.

29. H. Wioland, E. Lushi, R. E. Goldstein, Directed collective motion of bacteria under channel confinement. *New J. Phys.* **18** (2016), doi:10.1088/1367-2630/18/7/075002.

30. E. Lushi, H. Wioland, R. E. Goldstein, Fluid flows created by swimming bacteria drive self-organization in confined suspensions. *Proc. Natl. Acad. Sci.* . **111**, 9733–9738 (2014).

31. A. Doostmohammadi, T. N. Shendruk, K. Thijssen, J. M. Yeomans, Onset of meso-scale

turbulence in active nematics. *Nat. Commun.* **8**, 1–7 (2017).

32. V. de Lorenzo, Synthetic microbiology: from analogy to methodology. *Microb. Biotechnol.* **10**, 2–4 (2017).

33. A. Be'er, G. Ariel, O. Kalisman, Y. Helman, A. Sirota-Madi, H. P. Zhang, E.-L. E.-L. Florin, S. M. Payne, E. Ben-Jacob, H. L. Swinney, Lethal protein produced in response to competition between sibling bacterial colonies. *Proc. Natl. Acad. Sci.* **107**, 6258–6263 (2010).

34. A. Be'er, H. P. Zhang, E.-L. E.-L. Florin, S. M. Payne, E. Ben-Jacob, H. L. Swinney, Deadly competition between sibling bacterial colonies. *Proc. Natl. Acad. Sci. .* **106**, 428–33 (2009).

35. S. Kohlmeier, T. H. M. Smits, R. M. Ford, C. Keel, H. Harms, L. Y. Wick, Taking the fungal highway: Mobilization of pollutant-degrading bacteria by fungi. *Environ. Sci. Technol.* **39**, 4640–4646 (2005).

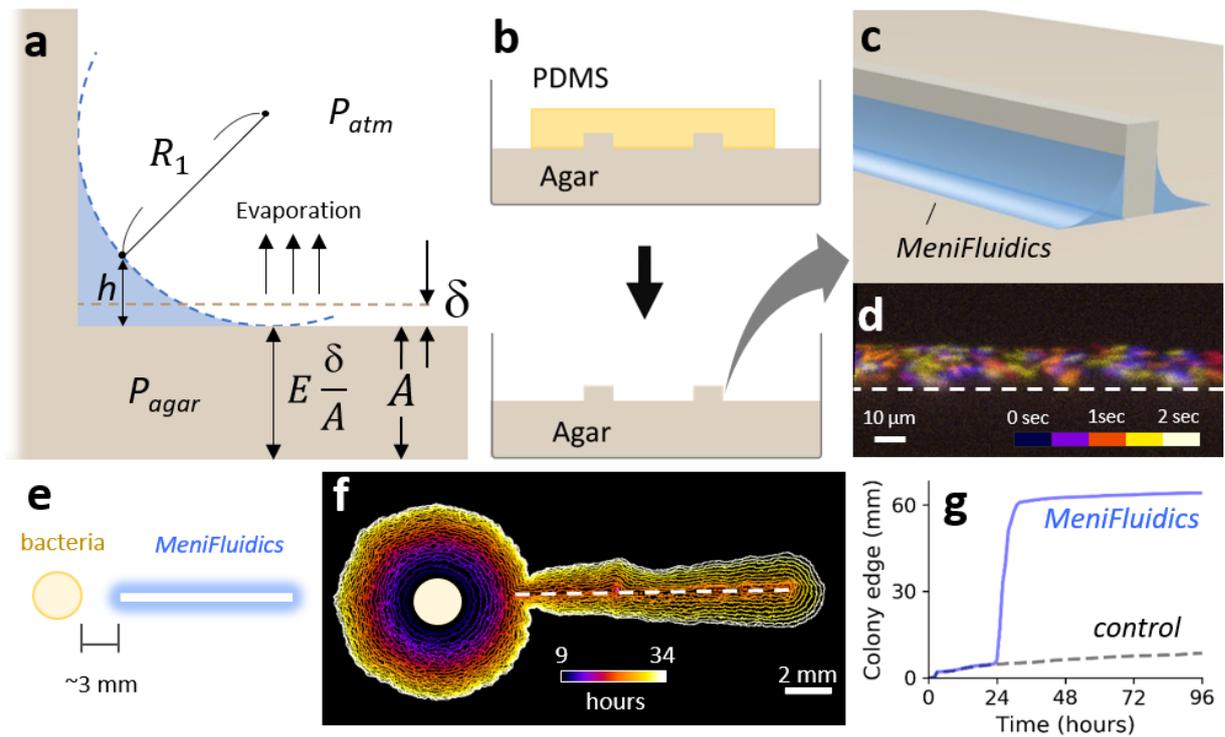

**Figure 1.** A) Schematic illustration of meniscus layer (blue) formed on structured agar surface (beige). When $1/R_2 \sim 0$, $R_1 = \sigma/E\Delta$ as denoted by the Young-Laplace equation (Eq.1), $h = R_1(1 - sin\frac{\pi}{4}) \approx 0.3R_1$, $E$ is Young's modulus, $\delta$ is the volume loss due to evaporation, $A$ is the volume of agar, $\sigma$ is surface tension, and $P$ is the pressure. B, C) Schematics of the MeniFluidics formation through PDMS moulding on agar. D) Projected time-lapse microscopy image of meniscus channel populated by bacteria. Colours denote time points of observation. Dashed line indicates the edge of structure (see Figure S3 for corresponding brightfield image). E) Schematic illustration of experiment where bacteria were inoculated ~3 mm away from MeniFluidics. F) Representative time evolution of a biofilm grown on agar with MeniFluidics. G) Distances of biofilm colony edge from the centre on flat agar surface (dashed grey) and MeniFluidics channel (blue).

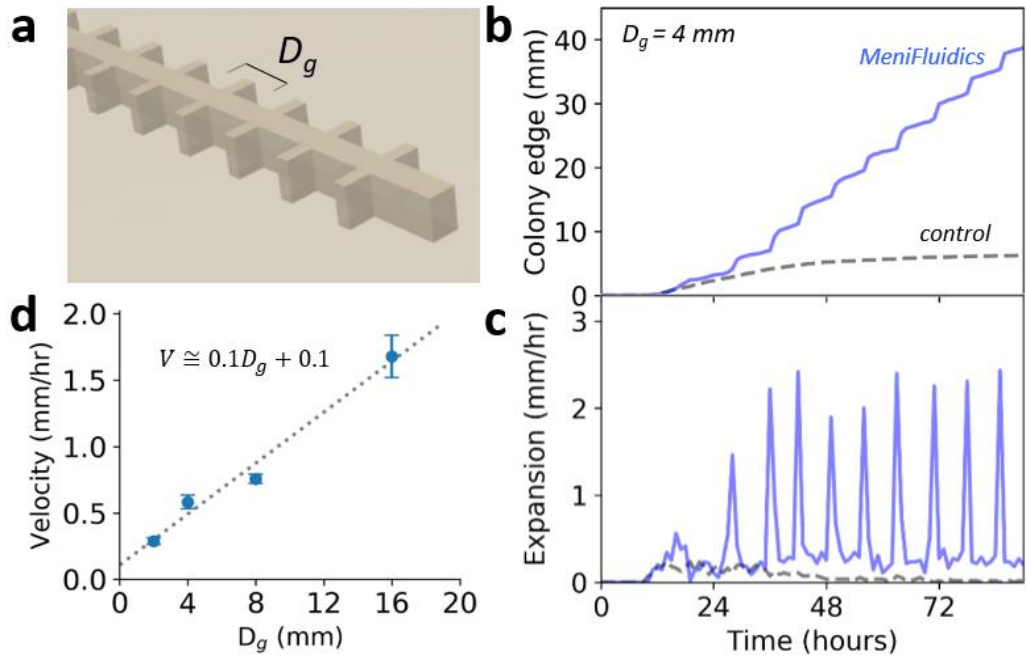

**Figure 2.** A) Schematic illustration of 3D structure with hinges. $D_g$ is the distance between hinges. B) Representative graphs of colony edge and C) expansion speed over time along hinged MeniFluidics with $D_g = 4\ mm$. D) Mean overall expansion speed as a function of $D_g$. Dashed line is a linear regression. Error bars are standard deviations.

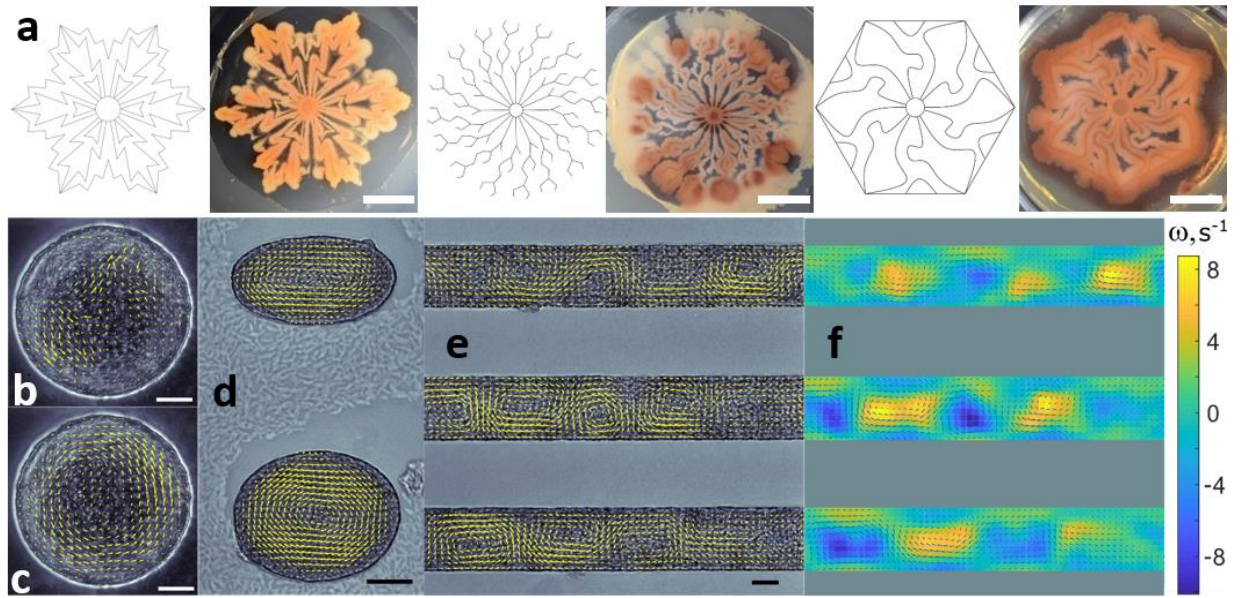

**Figure 3** A) Arbitrary designed MeniFluidics patterns (left) and patterned biofilms formed by the corresponding designs (right). Scale bar, 2 cm. B) Agar well populated by swarming bacteria revealing the chaotic flow of bacterial active fluid. C) The flow field, $\vec{V}$, averaged over 1000 instantaneous fields reveals a mean spiral flow. D) Averaged fields in the wells of elliptical shape represent a stable elongated vortex. E) Averaged velocity field in a long rectangular groove is organized into a vortex lattice. F) Vorticity field, $\omega = \nabla \times \vec{V}$, calculated from panel D. Scale bar, 20 $\mu m$ (B-F). Further details of this investigation will be published elsewhere.

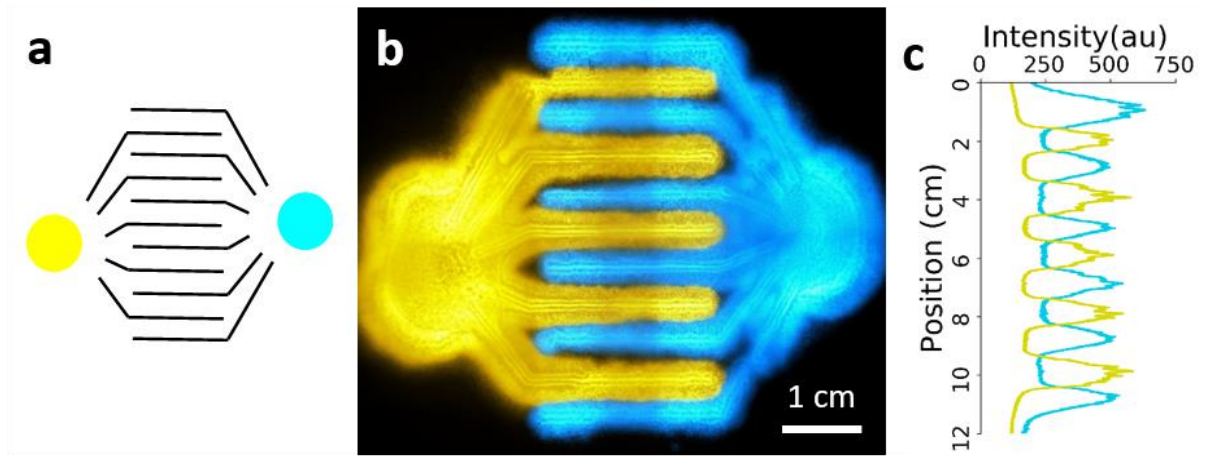

**Figure 4. Engineering of two bacterial strains into a pattern.**

A) Schematic of interdigitated MeniFluidic design. Yellow and cyan circles are inoculation points of fluorescently labelled bacterial strains. B) Representative fluorescence microscopy image of biofilms formed by YFP (yellow) or CFP (cyan) expressing bacteria. C) Fluorescence intensity across the dashed line in panel B.

# Supporting Materials

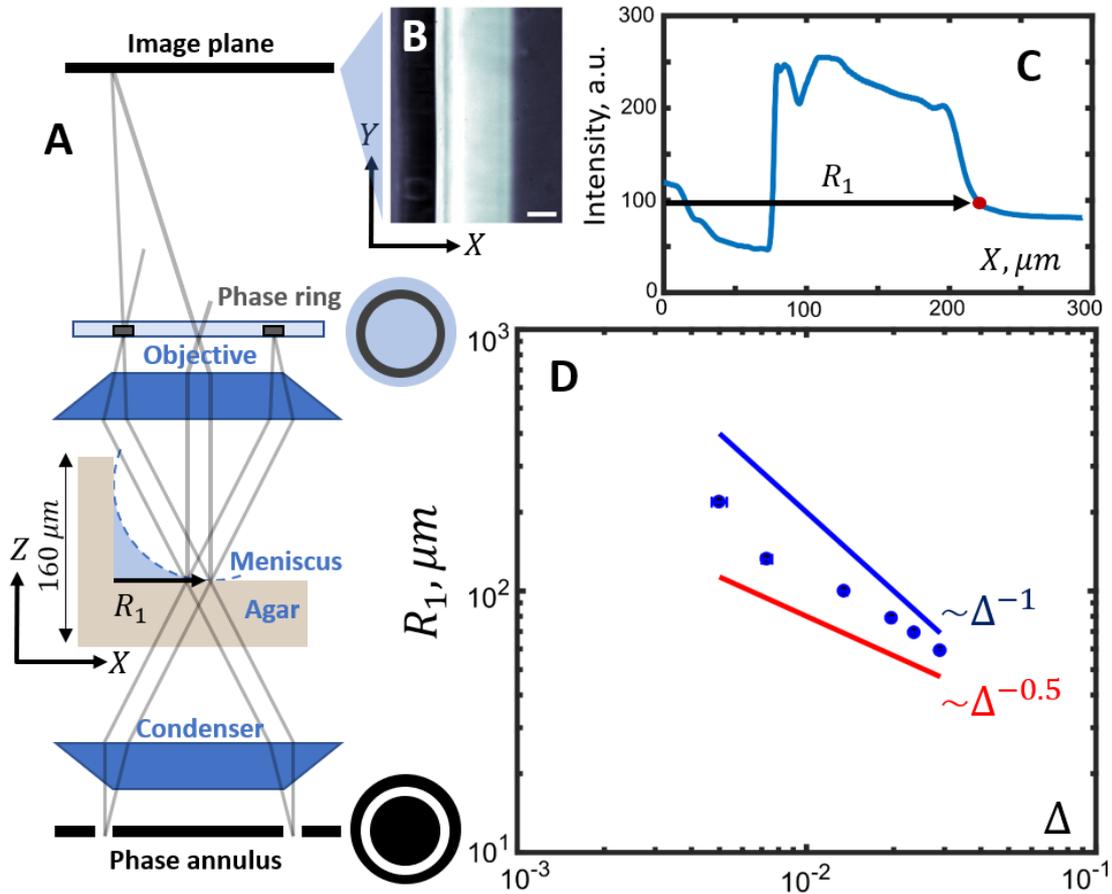

**Fig. S1.** A) Schematics of the phase contrast microscopy to detect the meniscus radius. B) Example of the meniscus image. Left side of the image is aligned with the inner corner of the structure. Scale bar is 50 $\mu m$. The initial dark strip is related to the optical shading from the tall structure. C) Intensity profile of the image to identify the tip of the meniscus. D) Dependence of the meniscus radius on the relative reduced mass of the agar gel.

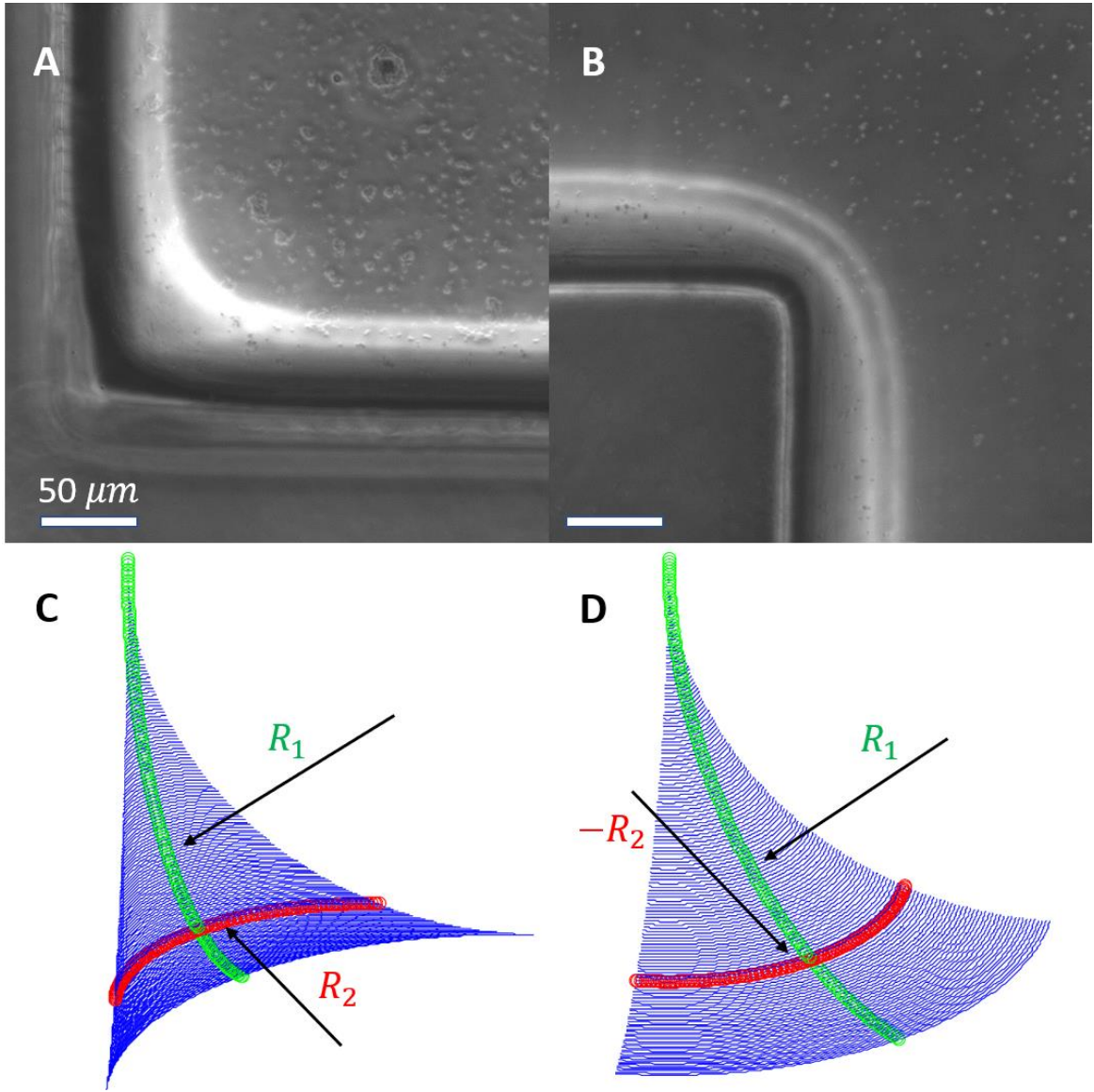

**Fig. S2.** A,B) Microscopy images of menisci on agar structure. Height of the structure is 160 microns in 1.5% agar, $\Delta$ = 0.01. Meniscus is bent at corners of the ridges, where a) is a concave corner, and b) is a convex corner. C,D) Diagrams showing two principal radii of curvature for a bent meniscus. C) corresponds A), while D) corresponds B).

**1.** Pour melted agar on solidified agar plate.

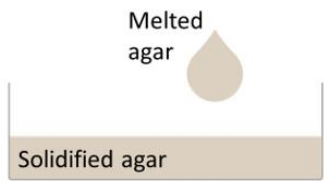 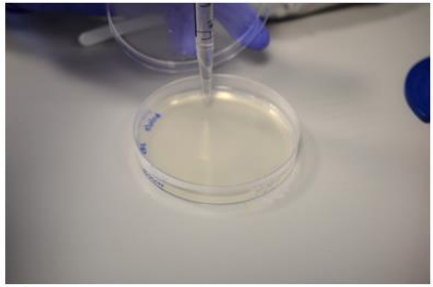

**2.** Immediately place PDMS mould.

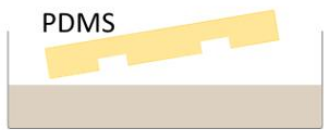 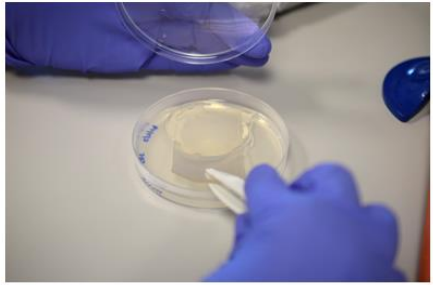

**3.** Remove air bubbles between agar and PDMS.

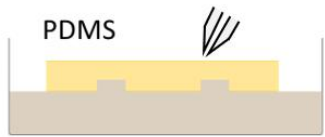 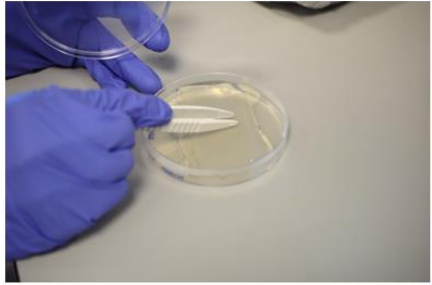

**4.** Remove PDMS.

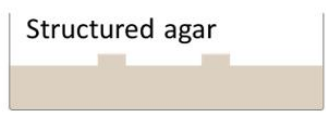

**Fig. S3.** Illustrative protocol of the micro patterning procedure.

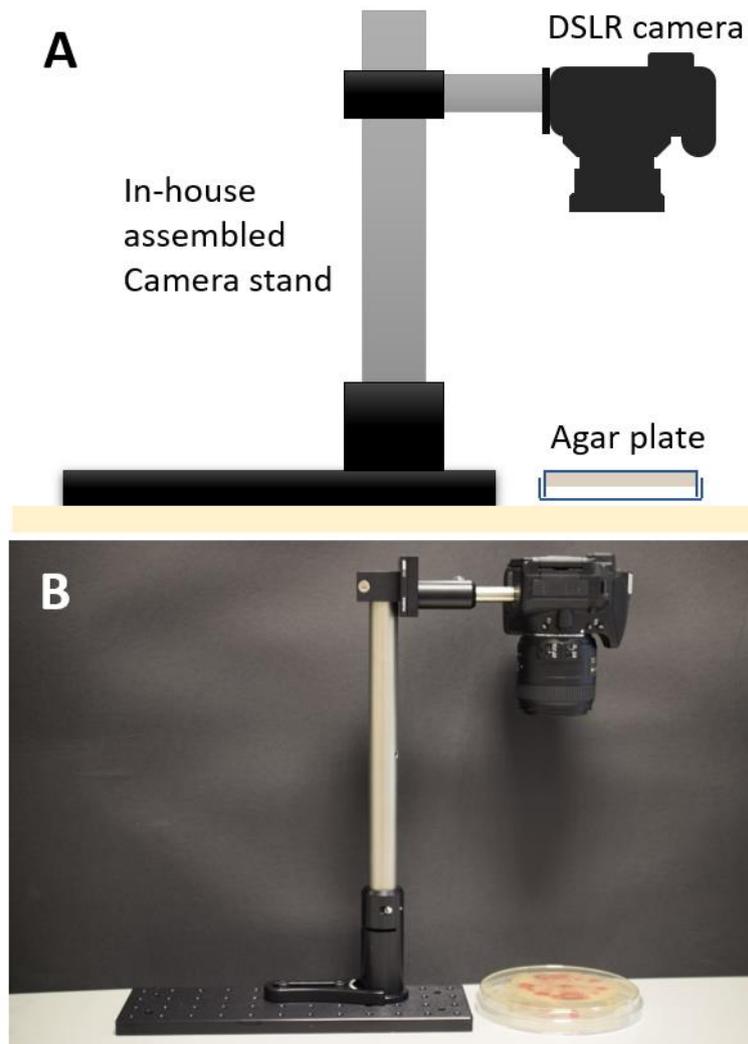

**Fig. S4.** DSLR camera imaging platform. A) Schematics and B) photograph of the imaging platform used in this study. The items to assemble this are in Table S2.

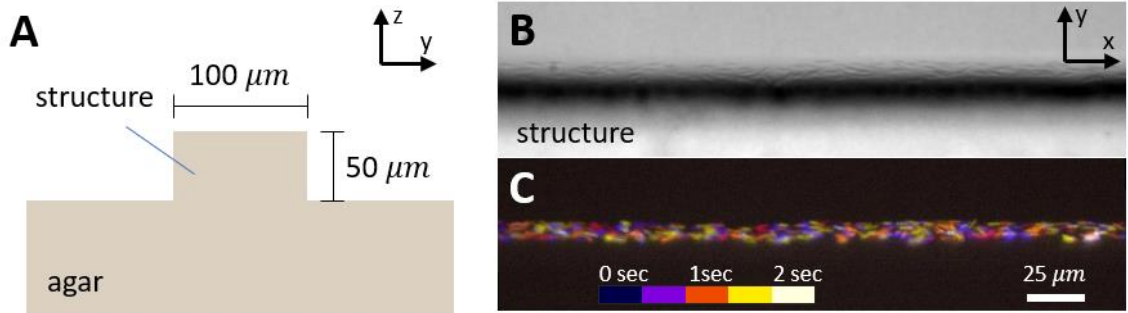

**Fig. S5.** Single-cell microscopy images along MeniFluidics. A) Schematic of structured agar surface B) phase contrast microscopy image of the structure. C) Combined fluorescence microscopy image of bacterial cells. Bacterial cells are pseudo-coloured for each frame. The colorbar represents the time scale. See Movie S1.

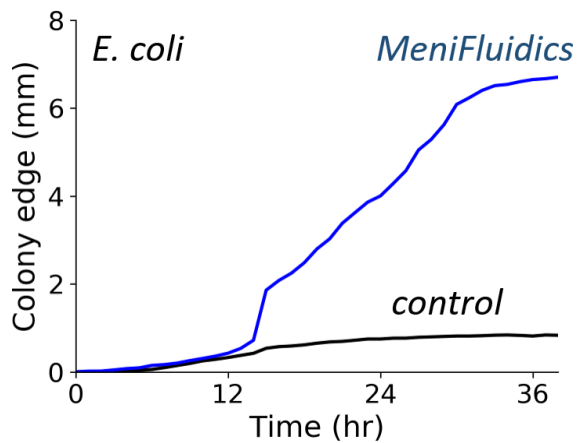

**Fig. S6.** E. coli colony expansion was accelerated by MeniFluidics.

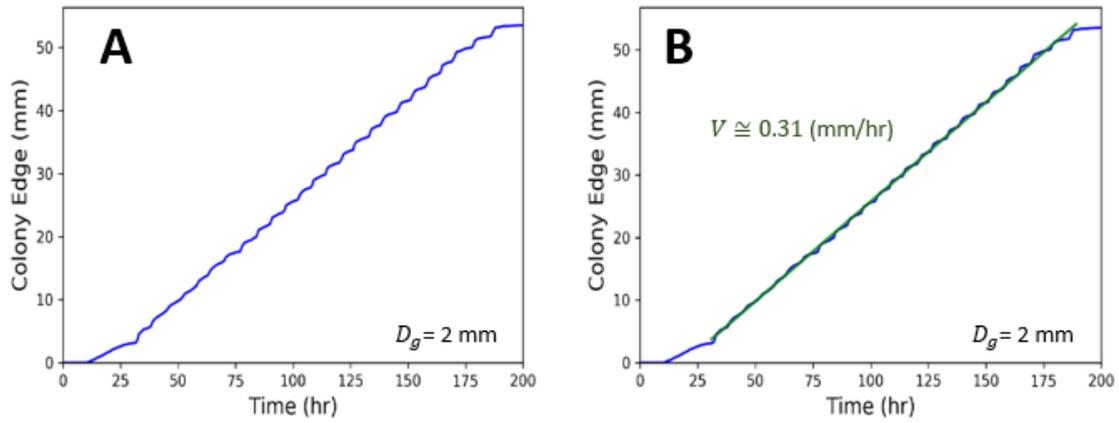

**Fig. S7.** A) Representative graph of colony edge expansion over time along hinged MeniFluidics channel. B) A linear fit to estimate the overall colony expansion speed.

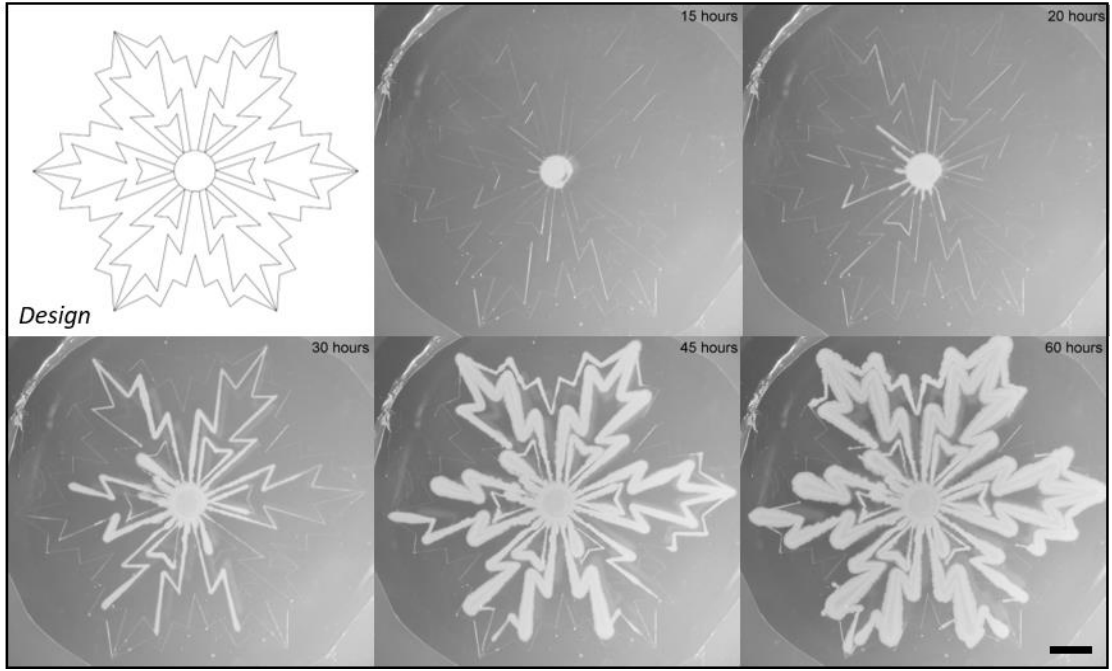

**Fig. S8.** Time-lapse sequence of arbitrary shaped biofilm colony made with MeniFluidics. Images correspond to Fig 3A. Scale bar is 1 cm.

**Table S1.** Bacterial strains used in the experiments.

| Strain | Genotype | Source |
|---|---|---|
| MA002 | *B. subtilis* NCIB 3610 wildtype | (Irnov and Winkler, 2010) |
| MA022 | *B. subtilis* NCIB 3610 AmyE::P$_{ftsZ}$-CFP, spec$^R$ | (Asally et al., 2012) |
| MA351 | *B. subtilis* NCIB 3610 AmyE::P$_{ftsZ}$-YFP, cm$^R$ | (Asally et al., 2012) |
| MA040 | *B. subtilis* NCIB 3610 ScaA::P$_{hag}$-YFP, cm$^R$ | (Asally et al., 2012) |

**Table S2.** Items for assembling imaging stand

| Item | Thorlabs catalogue number |
|---|---|
| Solid aluminium breadboard 100 mm x 300 mm x 12.7 mm, M6 | MB1030/M |
| Ø25.0 mm Pillar Post, M6 Taps, L = 300 mm | RS300/M |
| Clamping Fork for Ø1.5" Pedestal Post or Post Pedestal Base Adapter | PF175 |
| Post Mounting Clamp for Ø25 mm Post, Metric | C1001/M |
| Ø25 mm Post Holder with Flexure Lock, Pedestal Base, L = 100 mm | RSH4/M |
| Post | TR3 |
| Post holder | PH50/M |

**Movie S1 (separate file).** Time lapse of bacterial motion in meniscus layer.

**Movie S2 (separate file).** Biofilm expansion along MeniFluidics structure corresponding to Fig. 1E-G and S5.

**Movie S3 (separate file).** Biofilm expansion along MeniFluidics hinged structure with hinges corresponding the Fig. 2 and S7.

**Movie S4 (separate file).** Biofilm expansion along designed MeniFluidics structure corresponding the Fig. 3A and S8.

**Movie S5 (separate file).** Bacteria turbulence in a well, corresponding the Fig. 3B.

**Movie S6 (separate file).** Biofilm expansion along MeniFluidics designed by an obstacle (metal washer) placed on agar surface.